\documentclass{article}

\usepackage{microtype}
\usepackage{graphicx}
\usepackage{subfigure}
\usepackage{booktabs} 

\usepackage{hyperref}

\usepackage[accepted]{synsml2023}

\usepackage{amsmath}
\usepackage{amssymb}
\usepackage{mathtools}
\usepackage{amsthm}
\usepackage{epstopdf}
\usepackage[capitalize,noabbrev]{cleveref}

\theoremstyle{plain}

\theoremstyle{definition}

\theoremstyle{remark}

\usepackage[textsize=tiny]{todonotes}

\synsmltitlerunning{Submission and Formatting Instructions for Syns \& ML at ICML 2023}

\begin{document}

\twocolumn[
\synsmltitle{Diffusion model based data generation for partial differential equations}




\begin{synsmlauthorlist}
\synsmlauthor{Rucha Apte}{comp}
\synsmlauthor{Sheel Nidhan}{comp}
\synsmlauthor{Rishikesh Ranade}{comp}
\synsmlauthor{Jay Pathak}{comp}
\end{synsmlauthorlist}

\synsmlaffiliation{comp}{Ansys, Inc., Canonsburg, PA , USA}
\synsmlcorrespondingauthor{Rucha Apte}{rucha.apte@ansys.com}

\synsmlkeywords{Machine Learning}

\vskip 0.3in
]



\printAffiliationsAndNotice{}  

\begin{abstract}

In a preliminary attempt to address the problem of data scarcity in physics-based machine learning, we introduce a novel methodology for data generation in physics-based simulations. Our motivation is to overcome the limitations posed by the limited availability of numerical data. To achieve this, we leverage a diffusion model that allows us to generate synthetic data samples and test them for two canonical cases: (a) the steady 2-D Poisson equation, and (b) the forced unsteady 2-D Navier-Stokes (NS) {vorticity-transport} equation in a confined box. By comparing the generated data samples against outputs from classical solvers, we assess their accuracy and examine their adherence to the underlying physics laws. In this way, we emphasize the importance of not only satisfying visual and statistical comparisons with solver data but also ensuring the generated data's conformity to physics laws, thus enabling their effective utilization in downstream tasks.


\end{abstract}

\section{Introduction}

The application of {{machine learning (ML) and}} deep learning (DL) techniques in modeling partial differential equations (PDEs) has gained significant momentum over the past decade. These techniques have been employed to address various challenges in physics-based modeling, such as developing closure terms for large-eddy simulations and Reynolds-averaged Navier-Stokes equations in computational fluid dynamics (CFD) \citep{duraisamy2021perspectives, maulik2019subgrid, ling2016reynolds}, enhancing computational efficiency for classical solvers \citep{bar2019learning, weymouth2022data}, and facilitating reduced-order modeling \citep{murata2020nonlinear, eivazi2022towards}, among others. Additionally, the field of physics-based DL research has witnessed the emergence of numerous frameworks tailored for fast inference and generalizability across different classes of PDEs. Examples include physics-informed neural networks (PINNs) \citep{raissi2019physics}, Fourier neural operators (FNOs) \citep{li2020fourier}, DeepONet \citep{lu2019deeponet}, latent-space based local learning \citep{ranade2020discretizationnet, ranade2022composable} and more. {Although these methods have demonstrated remarkable success in solving PDEs for a variety of applications, one of the main factors impacting their accuracy and generalizability is the scarcity of high-quality training data \cite{sun2017revisiting}.

In recent years, denoising diffusion models have emerged as the leading technique for generative modeling \citep{sohl2015deep, song2020improved, ho2020denoising}. These models follow a two-step process: A forward step, where noise is added in a Markovian manner, followed by a reverse denoising step learnt using a deep neural network. Once trained, the model can generate new samples by starting from various realizations drawn from a Gaussian noise distribution. Diffusion models have demonstrated tremendous success in various domains, including conditional and unconditional image generation \citep{dhariwal2021diffusion, rombach2022high}, speech generation \citep{chen2020wavegrad}, image super-resolution \citep{saharia2022image}, video generation \citep{ho2022video}, to name a few. However, in the physics domain, the utilization of diffusion models has been relatively limited. \citet{shu2023physics} proposed a physics-inspired diffusion model for generating high-fidelity CFD data from low-fidelity/undersampled snapshots. \citet{vlassis2023denoising} used denoising diffusion models for conditional gneration of micsrostructures, testing it on the mechanical MNIST dataset. \citet{yang2023denoising} used a diffusion-based model for temporal prediction of a chaotic fluid flow. Without encoding prior physics constraint, they found that the diffusion-based network had comparable performance to that of existing models. \citet{lim2023score} extended the use of diffusion models to map functional spaces. Trained on a single resolution, the authors demonstrated the generation of PDE solutions for a variery of resolutions across different use cases.  

In this study, we use denoising diffusion implicit models (DDIMs) for the unconditional generation of data for two distinct physical systems: 2-D Poisson and 2-D Navier-Stokes flow. While our trained model lacks any prior physics encoding, we utilize physics-based constraints to select snapshots that adhere to fundamental laws. We propose two approaches to apply the physics-based constraint: PDE residual calculation (for the 2-D Poisson equation) and comparison with solver output (for the 2-D NS equation). Our aim is that, in the future, the data generation paradigm based on diffusion models can partially alleviate the challenge of data scarcity for physics-based machine learning models.

\section{Methodology}

\subsection{Diffusion Model and Architecture}

We utilize a diffusion model with a cosine scheduler that progressively degrades the data over 1000 steps \cite{sehwag_minimal}. The reverse diffusion process is parametrized by a deep neural network based on the widely used U-Net architecture in the diffusion literature. To facilitate efficient reverse sampling from Gaussian noise, we employ the DDIM strategy described in Song et al. (2020) \cite{song2020denoising} and use 500 sampling steps (instead of 1000) to accelerate the sampling speed. The input configuration of the U-Net architecture depends on the specific data to be modeled. For example, in case of the 2-D Poisson equation ($\nabla ^{2} u = f$), both variables $u$ and $f$ are provided as input to the U-Net through two separate channels. For problems involving temporal variation, different timesteps can be provided as inputs to the U-Net architecture through separate channels. At the time of sampling, all the channels are initialized with Gaussian noise and DDIM is used to arrive at $T = 0$ from $T = 1000$, thus obtaining denoised channels. 

\subsection{Physics-Based Constraints for Selection of Generated Data}

For physics-based ML data, it is crucial to ensure that the generated data samples, whether obtained from a traditional solver or a machine learning approach, adhere to the underlying governing equations in addition to being visually and statistically accurate. To verify this, we propose two distinct approaches. In the first approach, we compute the MSE of the PDE residual over the grid, for example $\mathrm{MSE}(|\nabla^2 u - f|)$ in the case of a 2-D Poisson's equation. We selectively retain only those samples where the $\mathrm{MSE}$ is less than a particular threshold. This criterion ensures that the the generated solutions satisfy the governing equation. In the second approach, one can use a traditional solver to verify the quality of the generated data. In the case of steady state PDEs, we compare the MSE between the generated solution ($u$ in Poisson's equation) with the $u$ evaluated from a classical solver for the same generated $f$. Alternatively, for transient PDEs, the first channel of the generated data is provided as an input to a traditional solver and the $\mathrm{MSE}$ (averaged across the entire grid and remaining channels) of the diffusion-generated data and solver-generated data is evaluated. We retain only those sets of generated snapshots that exhibit an MSE lower than a certain threshold. This approach guarantees that the selected samples align with the underlying physics.


\section{Experiments}

We demonstrate our diffusion model based data generation technique for two distinct use cases outlined below.

\subsection{2-D Poisson Equation}
In the case of the 2-D Poisson equation, $\nabla^2 u = f$, both $u$ and $f$ are passed as two separared channels to the U-Net architecture. The weights of the U-Net are optimized based on the loss function $L_{\mathrm{poisson}} = \lVert \epsilon_u - \epsilon_u^{\mathrm{pred}} \rVert^2 + \lambda \lVert \epsilon_f - \epsilon_f^{\mathrm{pred}} \rVert^2$, where $\epsilon$ corresponds to the noise added in the forward diffusion. We empirically found that $\lambda=2$ worked the best for 2-D Poisson equation.  The network is trained on {10,000} pairs of $[f,u]$ generated on a $64\times64$ grid using a multigrid solver. 

In this study, we aimed to address the challenge of generating f and u simultaneously using a diffusion model. Traditionally, data generation tasks often focus on generating one variable at a time, such as generating f or u independently. However, in our context, it was crucial to generate f and u together due to their inherent dependencies and interactions. This posed a more complex and challenging problem, as the diffusion model had to capture the joint distribution of f and u accurately.
\subsection{2-D Forced Navier Stokes Vorticity-Transport Equation}
For the forced unsteady 2-D Navier-Stokes (NS) equation, we train the diffusion model on blocks of five consecutive vorticity ($\omega$) fields, where separation between two consective fields in time {$\Delta t = 1.6\mathrm{s}$}. {The blocks size (five in this case) can be an arbitrary choice.} These five consecutive vorticity fields are sent as five channels input to the U-Net network, and the loss in the noise prediction is calculated by summing over all the five channels, $L_{\mathrm{NS}} = \sum_{c=1}^{5} \lVert \epsilon_c - \epsilon_c^{\mathrm{pred}} \rVert^{2}$, where $c$ is channel index. The network for 2-D NS equations is trained with {$700$} solutions, each starting with a different initial condition. The vorticity is evolved until $t = 320\mathrm{s}$ on $64 \times 64$ grid using a NS solver \cite{li2020fourier}. The viscosity is set at {$\nu = 10^{-4}$} and the forcing function takes the form $f = 0.1\mathrm{sin}(4\pi(x+y)) + 0.1\mathrm{cos}(4\pi(x+y))$. To increase the amount of data for training, a sliding window approach with a stride of three was used. Hence, there is an overlap between two consecutive blocks of five snapshots.

The U-Net architecture used for {2-D Poisson and NS equation data} consists of {$6M$ parameters} with adaptive group normalization. The codebase for this work is built on \cite{sehwag_minimal}.  The U-Net network is conditioned with the diffusion time $t$ through feature vector embedding. Both networks are trained for 250 epochs.

\section{Results}

\subsection{2-D Poisson Equation}

\begin{figure}[ht]
\vskip 0.2in
\begin{center}
\centerline{\includegraphics[width=1\columnwidth]{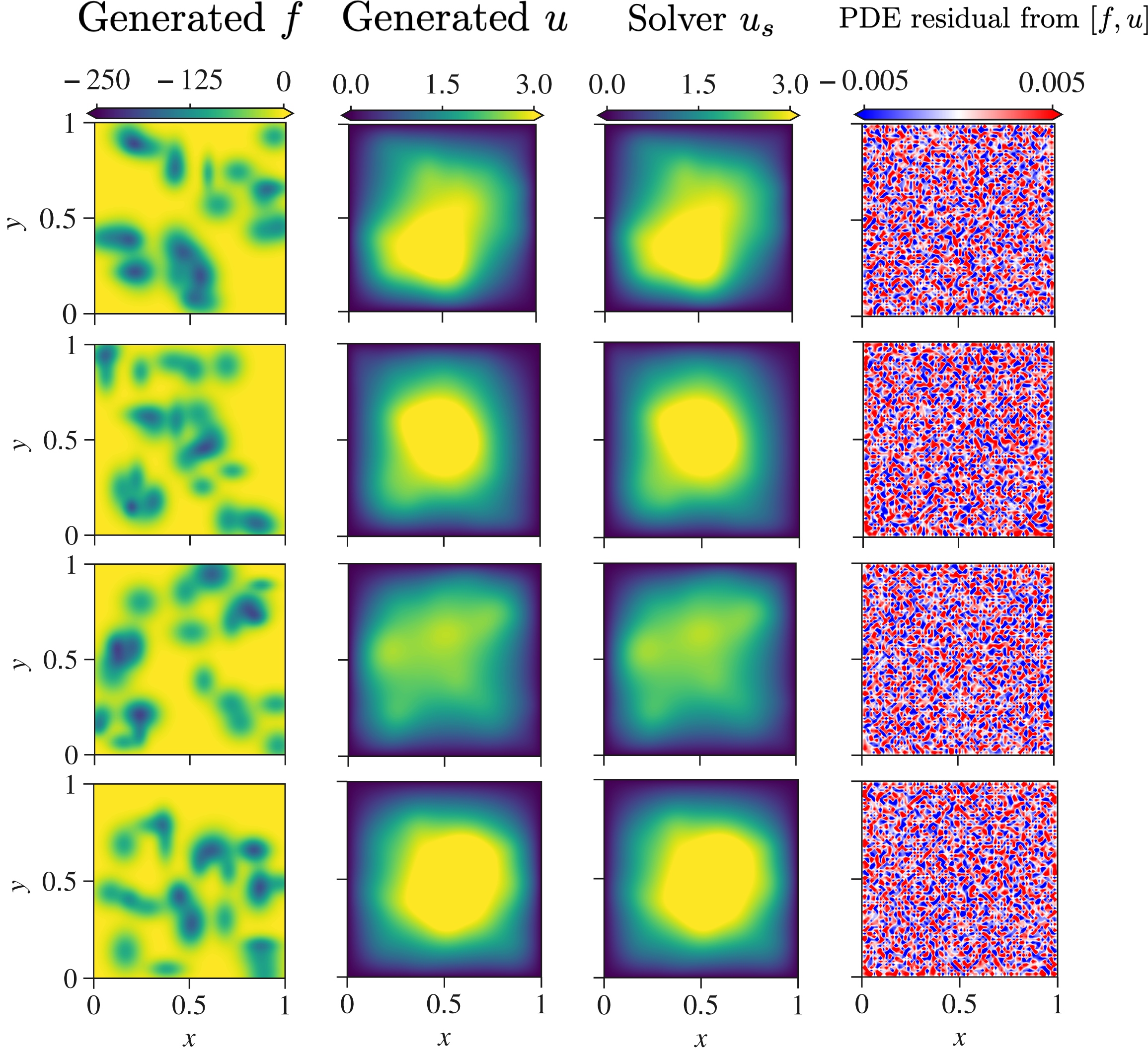}}
\caption{Four $[f_{\mathrm{generated}},u_{\mathrm{generated}}]$ pairs generated from diffusion model (first two columns from left), $u_{\mathrm{solver}}$ generated from the corresponding $f$ (third column), PDE residual calculated using generated $[f,u]$ pairs (fourth column). $\mathrm{MSE}$ of PDE residuals for all these four generated samples  $<6\times 10^{-6}$.}
\label{poisson2d}
\end{center}
\vskip -0.2in
\end{figure}

In the context of 2-D Poisson, we conducted a comprehensive analysis of the generated $[f, u]$ pairs, focusing on their visual quality and adherence to the underlying governing equations. Figure \ref{poisson2d} showcases four such pairs (left two columns). The contours of the synthetically generated $[f, u]$ exhibit smooth boundaries, and their values fall within the expected range. However, it is important to note that visual appearance alone does not guarantee adherence to the underlying physics equations.

To ensure the physical validity of the generated images, we examine the (PDE) residual, $\nabla^{2} u_{\mathrm{generated}} - f_{\mathrm{generated}}$, associated with each generated pair (last column in Figure \ref{poisson2d}). All the samples in the figure correspond to $\mathrm{MSE}$ of PDE residual $< 6 \times 10^{-6}$.
When comparing them to the solver-generated $u_{\mathrm{generated}}$ (third column), obtained by passing $f_{\mathrm{generated}}$ through a finite-difference solver, we observe that all the generated samples exhibit a very close visual resemblance between $u_{\mathrm{generated}}$ and $u_{\mathrm{solver}}$. 

\begin{table}[t]
\caption{Relative $L_2$ error between the mean and standard deviation of the solver-generated data distribution and the diffusion-generated data distribution for 2-D Poisson equation. We condition the diffusion statistics for 2-D Poisson on $\mathrm{MSE}$ of residual $< 6\times 10^{-6}$.}
\label{poisson2d-table}
\vskip 0.15in
\begin{center}
\begin{small}
\begin{sc}
\begin{tabular}{lcccr}
\toprule
Data & Mean & Standard deviation  \\
\midrule
2-D Poisson, $f$    & {$0.007$} & {$0.007$} \\
2-D Poisson, $u$    & {$0.004$} & {$0.01$}  
\\
\bottomrule
\end{tabular}
\end{sc}
\end{small}
\end{center}
\vskip -0.1in
\end{table}

Table \ref{poisson2d-table} shows the relative $L_2$ error between the statistics of synthetic and solver generated $f$ and $u$. For both variables, the percentage error is less than 1\% , indicating the efficacy of a diffusion-based model in recovering the statistics of the data distribution. For the threshold of $\mathrm{MSE}$ of PDE residual $< 6\times 10^{-6}$, $99962$ pairs out of total generated $100,000$ pairs were admissible and can be used for other downstream tasks. 

Figure \ref{poisson2d} and table \ref{poisson2d-table} clearly demonstrates that generated data with low PDE residuals demonstrate a very close agreement with the underlying physics equations and exhibit a greater visual  and statistical resemblance to solver solutions, making them more suitable for subsequent analysis and utilization. Conversely, generated data with high PDE residuals should be approached with caution, as they may deviate significantly from the desired physical behavior. 

\subsection{2-D Forced Navier Stokes Vorticity-Transport Equation}

\begin{figure}[ht]
\vskip 0.2in
\begin{center}
\centerline{\includegraphics[width=1.1\columnwidth]{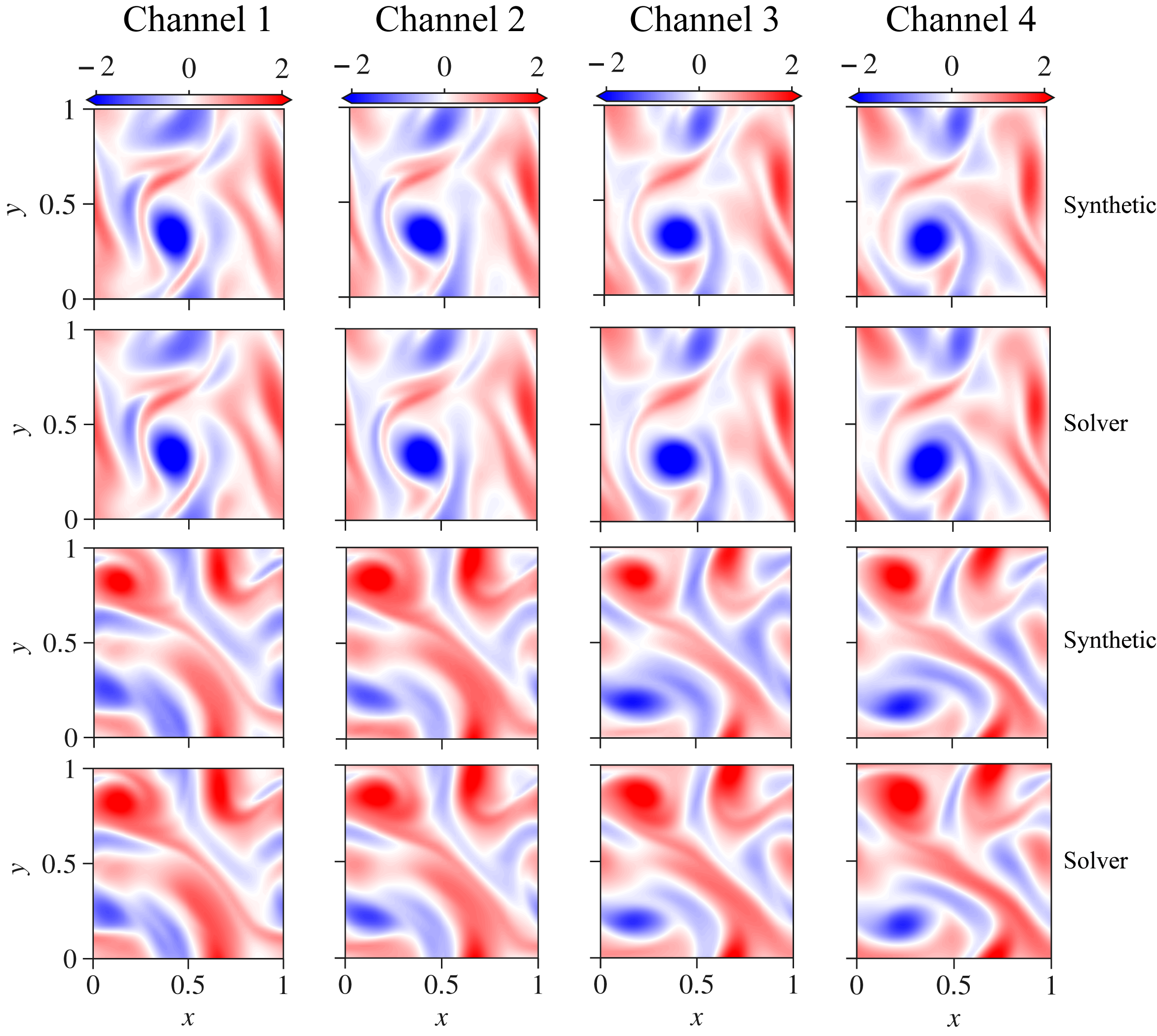}}
\caption{Diffusion- and solver-generated vorticity snapshots for 2-D forced NS system. Top two rows correspond to $\mathrm{MSE}$ $\approx 8 \times 10^{-3}$ between solver-generated and diffusion-generated snapshots. Bottom two rows correspond to $\mathrm{MSE} \approx 0.045$. }
\label{fno2d}
\end{center}
\vskip -0.2in
\end{figure}

{Figure \ref{fno2d} (first and third rows) shows the evolution of diffusion-generated vorticity. Corresponding to the first and third rows, we present the solver-generated vorticity in the second and fourth rows, respectively. The colorbar highlights the magnitudes, confirming that they fall within the expected range. Unlike the analysis performed on 2-D Poisson data, which involved evaluating PDE residuals, we adopt a different approach to validate the physical correctness of the data generated on 2-D NS vorticity-transport equation. The first snapshot is passed as an initial condition to the solver. Hence, only the next four snapshots are compared in figure \ref{fno2d}. The top two rows correspond to $\mathrm{MSE} \approx 8 \times 10^{-3}$ between diffusion-generated and solver-generated snapshots. We find that the diffusion-generated snapshots qualitatively capture the flow dynamics -- note the roll-up of vortex around $[x,y] = [0.5, 0.5]$ in both diffusion- and solver- generated snapshots (top two rows). For bottom two rows, although the flow dynamics between diffusion and solver generated snapshots look qualitatively consistent, $\mathrm{MSE}$ (calculated over the grid and the four snapshots) is one order of magnitude higher, at $0.045$. 

\begin{figure}[ht]
\vskip 0.2in
\begin{center}
\centerline{\includegraphics[width=\columnwidth]{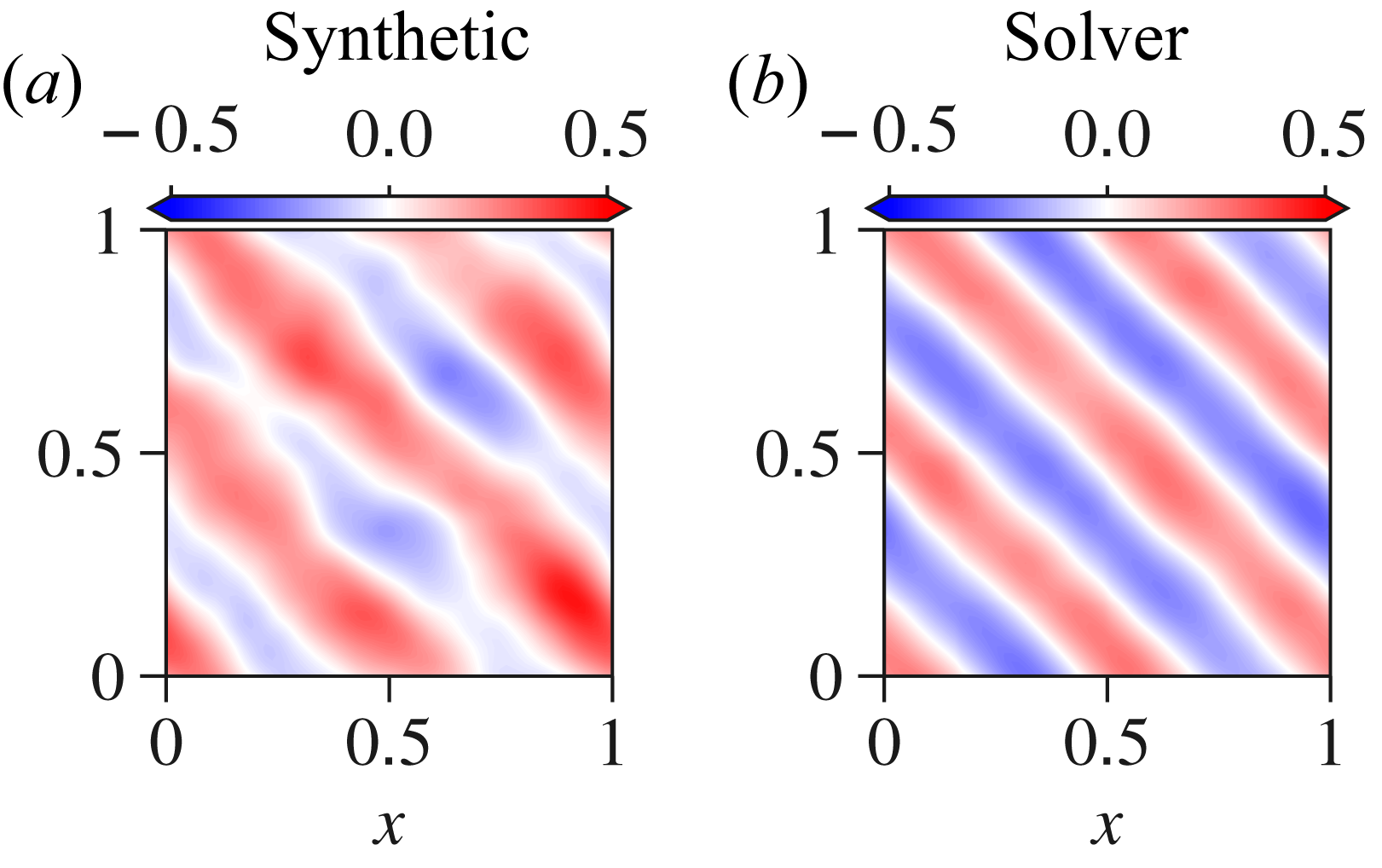}}
\caption{Two-dimensional contours of the mean vorticity field from the (a) diffusion-generated data distribution and (b) solver-generated distribution. For statistics on the diffusion-generated vorticity field, we condition on $\mathrm{MSE}$ between diffusion-generated and solver-generated snapshots $<2\times 10^{-2}$.}
\label{stats_fno}
\end{center}
\vskip -0.2in
\end{figure}

Figure \ref{stats_fno} shows the mean distribution of solver-generated vorticity field and diffusion-generated vorticity field. The solver-generated mean field (figure \ref{stats_fno}b) show inclined patches of alternate sign vorticity, reminiscent of the forcing function that is being applied. We find that the diffusion model is able to reproduce the inclined patches, corresponding to the forcing pattern qualitatively (figure \ref{stats_fno}a). However, it is important to note that the distribution of positive vorticity patches appears to be more dominant in the generated data compared to the solver-generated distribution.

Finally, in addition to quantitative evaluation metrics, we conducted a visual inspection of the generated samples to assess their diversity for both datasets (attached in appendix). It was evident that the generated data samples exhibited a wide range of variations and distinct features for both 2-D Poisson and 2-D NS equations. The visually diverse nature of the generated samples indicates the effectiveness of diffusion model model in producing novel and unique outputs.}

\section{Conclusions}

In this work, we introduced a data generation methodology based on diffusion models and validated it for two canonical physical systems: 2-D Poisson and 2-D forced Navier-Stokes vorticity-transport equation. Our findings demonstrate that the diffusion model can effectively generate visually and statistically consistent samples. To leverage these samples for subsequent downstream tasks such as training physics-based machine learning algorithms, one can employ PDE-based residuals or solver-based filtering methods to select physically consistent samples. This approach ensures that the generated data adheres to the underlying physics and can be reliably used in further analyses.

It is important to note that this work is ongoing, and there are several future directions to explore. One future direction involves incorporating physics-based losses in the training and sampling algorithm of the diffusion model itself. {In our current model, data generation is limited to a specific resolution. However, we are concurrently working on implementing super-resolution techniques for diffusion models. This will enable interpolation between resolutions and the generation of high-fidelity images. We are also exploring the extension of this method to solutions represented on unstructured meshes. Additionally, we plan to utilize the governing parameters of the data to condition the model in the future, enhancing its general-purpose capabilities.} 


\section*{Broader Impact}

This research direction can have significant impact on various fields important to society, which also face the issue of data scarcity, e.g., climate science, material science, etc. Diffusion models can contribute to the advancement of these fields by providing a means to generate realistic and physically accurate data for training and validating machine learning algorithms. However, the reliance on machine learning techniques for physics-based simulations may raise concerns about the interpretability and explainability of the models, given that physics-based simulations are crucial to ensure safety in various domains, e.g., aeronautics, nuclear power, electronic appliances, etc.





\bibliography{example_paper}
\bibliographystyle{synsml2023}

\newpage
\appendix
\onecolumn
\section{Diversity of generated data for both 2-D Poisson and 2-D NS}

\begin{figure}[ht]
\vskip 0.2in
\begin{center}
\centerline{\includegraphics[width=1\columnwidth]{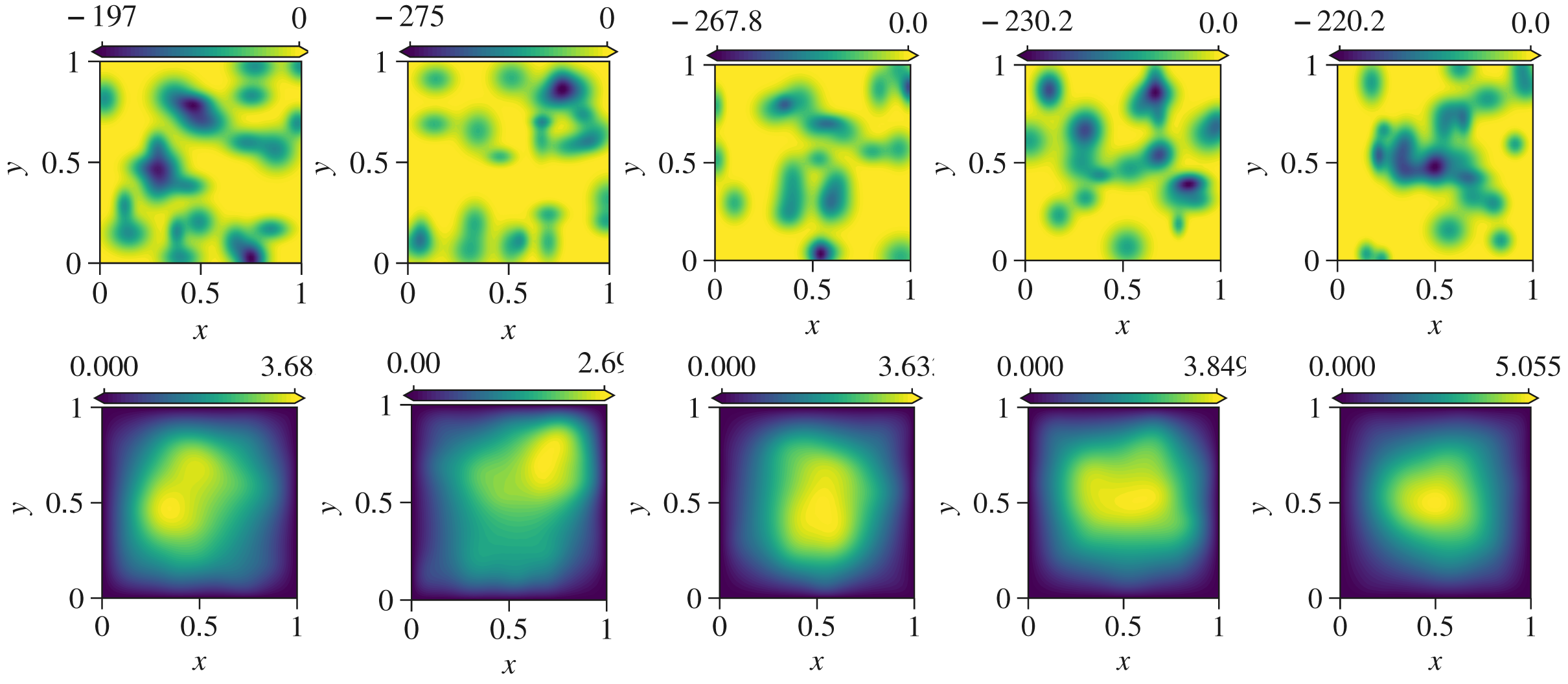}}
\caption{Example set of five randomly chosen $[f,u]$ pairs -- $f$ in top row and corresponding $u$ in bottom row for 2-D Poisson equation.}
\label{fno2d}
\end{center}
\vskip -0.2in
\end{figure}

\begin{figure}[ht]
\vskip 0.2in
\begin{center}
\centerline{\includegraphics[width=1\columnwidth]{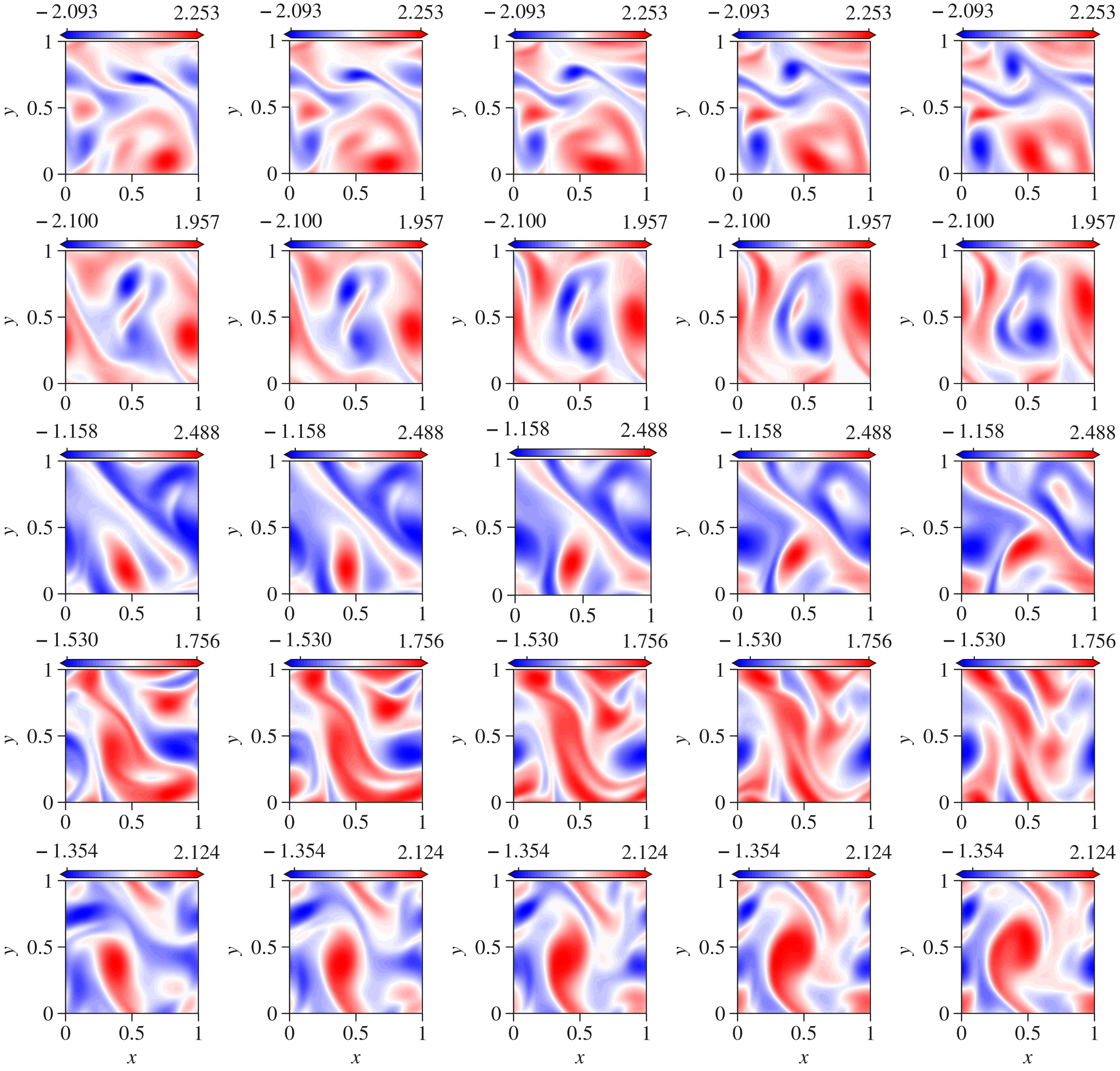}}
\caption{Example set of five randomly chosen vorticity fields for force 2-D NS equation. Snapshot 1 to 5 from leftmost to rightmost columns.}
\label{fno2d}
\end{center}
\vskip -0.2in
\end{figure}


\end{document}